\documentclass[letterpaper,12pt]{article}
\usepackage{amsmath,amssymb,amsfonts}
\usepackage{graphicx} 
\usepackage{empheq}
\usepackage{subcaption} 
\usepackage[section]{placeins} 
\usepackage{booktabs} 
\usepackage{algorithm} 
\usepackage{algpseudocode}
\usepackage{tikz-cd}
\usepackage{adjustbox} 
\usepackage{siunitx} 
\usepackage{natbib}

\DeclarePairedDelimiter\floor{\lfloor}{\rfloor}
\DeclareMathOperator{\sign}{\mbox{sign}}
\newcommand{\med}{\mbox{median}}
\newcommand{\RD}{\mbox{RD}}
\newcommand{\ls}{\leqslant}
\newcommand{\gs}{\geqslant}
\newcommand{\eps}{\varepsilon}

\newcommand{\hmu}{\hat{\mu}}
\newcommand{\hsigma}{\hat{\sigma}}

\newcommand{\ba}{\boldsymbol a}
\newcommand{\bb}{\boldsymbol b}
\newcommand{\bu}{\boldsymbol u}
\newcommand{\bv}{\boldsymbol v}
\newcommand{\bx}{\boldsymbol x}
\newcommand{\by}{\boldsymbol y}
\newcommand{\bz}{\boldsymbol z}
\newcommand{\bdelta}{\boldsymbol \delta}
\newcommand{\bmu}{\boldsymbol \mu}

\newcommand{\bA}{\boldsymbol A}
\newcommand{\bB}{\boldsymbol B}
\newcommand{\bC}{\boldsymbol C}
\newcommand{\bD}{\boldsymbol D}
\newcommand{\bI}{\boldsymbol I}
\newcommand{\bL}{\boldsymbol L}
\newcommand{\bS}{\boldsymbol S}
\newcommand{\bT}{\boldsymbol T}
\newcommand{\bV}{\boldsymbol V}
\newcommand{\bX}{\boldsymbol X}

\newcommand{\bZ}{\boldsymbol Z}
\newcommand{\bLambda}{\boldsymbol \Lambda}
\newcommand{\bSigma}{\boldsymbol \Sigma}

\newcommand{\btD}{\boldsymbol{\tilde{D}}}
\newcommand{\btS}{\boldsymbol{\tilde{S}}}
\newcommand{\btZ}{\boldsymbol{\tilde{Z}}}

\newcommand{\hbmu}{\hat{\boldsymbol \mu}}
\newcommand{\hbSigma}{\hat{\boldsymbol \Sigma}}

\begin{document}

\title{\bf Real-time outlier detection for large 
       datasets by RT-DetMCD}

\author{Bart De Ketelaere, Mia Hubert, 
        Jakob Raymaekers,\\ 
				Peter J. Rousseeuw, Iwein Vranckx\\
				\\
        KU Leuven, BE-3001 Heverlee, Belgium}
				
\date{January 25, 2020}

\maketitle

\begin{abstract}
Modern industrial machines can generate gigabytes 
of data in seconds, frequently pushing the 
boundaries of available computing power. 
Together with the time criticality of industrial 
processing this presents a challenging problem for 
any data analytics procedure. 
We focus on the deterministic minimum covariance
determinant method (DetMCD), which detects outliers
by fitting a robust covariance matrix.
We construct a much faster version of DetMCD by 
replacing its initial estimators by two new methods
and incorporating update-based concentration steps.
The computation time is reduced further by parallel
computing, with a novel robust aggregation method 
to combine the results from the threads.
The speed and accuracy of the proposed real-time 
DetMCD method (RT-DetMCD) are illustrated by 
simulation and a real industrial application to 
food sorting.
\end{abstract}


\section{Introduction}
Modern industries are data-rich environments 
where information from multiple sensors is 
captured at a high sampling frequency. 
Processing such data has to cope with typical 
challenges such as the presence of outliers. 
While classical statistical estimators can be 
highly affected by outliers, their robust 
counterparts can cope with a significant 
fraction of contamination. 
There is a vast literature about robust 
statistical techniques (e.g. 
\cite{daszykowski2007robust, 
Hubert:WIRE-MCD2, 
Rousseeuw:CritReviews,
Rousseeuw:Robreg}).
Although substantial research has already gone 
into constructing fast robust algorithms, more 
work is needed to be able to handle real-time 
multivariate situations with many thousands of 
observations per second, as required by some 
industrial processes. 

For this task we will focus on the Minimum 
Covariance Determinant (MCD) 
approach~\cite{Rousseeuw:LMS,Rousseeuw:MVE,
Hubert:WIRE-MCD2} which provides highly robust 
estimators for multivariate location and 
covariance matrices.
Its first practical algorithm was FastMCD 
\cite{Rousseeuw:FastMCD}.
More recently the DetMCD algorithm 
\cite{Hubert:DetMCD} was constructed, which
is deterministic unlike the random sampling
component of FastMCD.  
Although DetMCD is significantly faster
it is still prohibitive for the huge sample 
sizes envisaged here.
For routine use in real-time industrial 
environments we need to speed it up 
further, which motivated this research.

A recent review paper \cite{zhu2018review}
discussed the perspectives of robust methods 
for industrial process management when 
outliers are present.  
It highlighted several paths that can be
explored.
One of these is the evolution from a 
centralized analysis of large datasets towards 
parallel computing, whereby multiple threads 
work in parallel on data subsets after 
which the results are combined for the final 
result.  
Our work on DetMCD will indeed incorporate 
parallel computing.

The remainder of the paper is organized as follows. 
In Section~\ref{sec:DetMCD} we describe the DetMCD 
estimator and its main properties. 
Section~\ref{sec:serial} proposes an improved 
serial version which incorporates various new 
techniques and is substantially faster. 
Section~\ref{sec:parallel} constructs a parallelized
version, which speeds up computation even more.
The simulation in Section \ref{sec:sim} confirms the 
robustness, speed and accuracy of the proposed method.
Section \ref{sec:industrialexample} analyzes a real 
industrial dataset, and Section \ref{sec:concl}
concludes. 

\section{The Minimum Covariance Determinant approach} 
\label{sec:DetMCD}
Our goal is to detect outliers in a multivariate 
dataset with $n$ observations and $p$ variables. 
We denote the data by 
$\bX = (\bx_1,\ldots,\bx_n)^T$
where each observation 
$\bx_i = (x_{i1},x_{i2},\ldots,x_{ip})^T$ is a 
$p$-dimensional column vector. 
Here we assume that $p$ is moderate, say no more
than 40, otherwise a dimension reduction technique
such as robust PCA \cite{Hubert:ROBPCA} can 
be used.
The sample size $n$ should be higher than $p$ 
and is allowed to be huge, even up to several 
millions. 
We assume that the inliers roughly follow
a multivariate Gaussian distribution
$N(\bmu, \bSigma)$ with center $\bmu$ and 
covariance matrix $\bSigma$, possibly after 
transforming some skewed variables.

\subsection{The MCD estimator}
\label{sec:FastMCD}
Robust statistical methods aim to model
the inlying cases and then flag outliers as 
those observations that deviate too much from 
that model. 
Here we will focus on the Minimum 
Covariance Determinant (MCD) estimator
\cite{Rousseeuw:MVE}. 
Given a user-specified tuning constant $h$, 
where $[(n+p+1)/2] \ls h < n$, the raw MCD 
estimator is $(\hbmu_{raw},\hbSigma_{raw})$
where the location estimate $\hbmu_{raw}$ is 
the mean of the $h$ observations whose sample 
covariance matrix has the smallest 
determinant.
Intuitively these $h$ observations are
the most concentrated, since the determinant 
of a covariance matrix corresponds to the 
volume of its tolerance ellipsoid.
The scatter matrix estimate $\hbSigma_{raw}$ 
is that covariance matrix 
multiplied by the consistency factor 
$c(\alpha)$ of \cite{Croux:IFMCD} 
that depends on $\alpha = h/n$ and 
compensates for the fact that only $h$ 
out of $n$ observations are included.

The indices $i$ of these $h$ observations 
form a set $H$, called an $h$-subset. 
The raw MCD estimates are then given by
\begin{align}
	\hbmu_{raw} &= \frac{1}{h}
	 \sum_{i \,\textrm{in}\, H}\bx_i \;,\\
	\hbSigma_{raw} &= \frac{c(\alpha)}{h-1}\,
	\sum_{i \,\textrm{in}\, H} (\bx_i-\hbmu_{raw})
	(\bx_i-\hbmu_{raw})^T.
\label{eq:mcdraw}
\end{align}
Note that the MCD is only defined when 
$h > p$, otherwise the covariance matrix of 
any $h$-subset is singular, so we want 
$n > 2p$. 
In practice it is however recommended 
that $n$ be much larger, in order to
obtain a more accurate result. 

The raw MCD estimator is highly robust as it 
can withstand up to $n-h$ outliers. 
The breakdown value of an estimator is the 
proportion of outliers that can be resisted.
The breakdown value of the MCD is $1-\alpha$.
Choosing  $\alpha = 0.5$ yields an estimator 
with a maximal breakdown value of $50\%$ but 
a rather low statistical efficiency, whereas 
taking $\alpha = 0.75$ yields a more efficient 
estimator with lower $25\%$ breakdown value.

To increase the efficiency we carry out a 
reweighting step. 
For this we first measure how much each data 
point $\bx_i$ deviates from the raw MCD fit,
by computing the \textit{robust distances} 
$\RD_i=d(\bx_i,\hbmu_{raw},\hbSigma_{raw})$ 
where the statistical distance $d$ is 
defined as 
\begin{equation*}
	d(\bx,\bmu,\bSigma) = \sqrt{(\bx-\bmu)^T
	    \bSigma^{-1} (\bx-\bmu)} \;.
\label{eq:statdist}
\end{equation*}
The reweighted MCD estimates 
$(\hbmu_{rew},\hbSigma_{rew})$ are then
computed as the mean and covariance matrix 
of the observations $\bx_i$ 
whose $\RD_i$ do not exceed the cut-off 
value $c_p= \sqrt{\chi^2_{p,0.975}}$ (where
$\chi^2_p$ is the chi-squared distribution 
with $p$ degrees of freedom). 
Then outliers are flagged as those cases 
whose final robust distance 
$\RD_i = d(\bx_i,\hbmu_{rew},\hbSigma_{rew})$ 
exceeds $c_p$.
Note that a higher cutoff such as 
$\sqrt{\chi^2_{p,0.99}}$ could be chosen,
but in this paper the 0.975 quantile was
used throughout to be able to detect outliers
that are relatively close to the majority.
This was important in the application
on food sorting in Section 
\ref{sec:industrialexample},
where letting pass some foreign material 
creates bigger problems (such as regulatory) 
than discarding a small fraction of 
potentially clean food.

Note that the reweighted MCD inherits the 
breakdown value of the raw MCD, so setting
$\alpha = 0.5$ yields a reweighted estimator
with a breakdown value of 50\%.

When any nonsingular affine transformation is 
applied to the data (such as a rotation, 
a reflection or rescaling) the MCD estimator 
transforms along with it. 
This is called affine equivariance. 
Therefore the robust distances $\RD_i$
remain invariant under such a transformation.

The exact raw MCD is very hard to 
compute, as it requires the evaluation of all 
$\binom{n}{h}$ subsets of size $h$ which is 
infeasible for increasing $n$. 
The FastMCD algorithm of \cite{Rousseeuw:FastMCD}
approximates the MCD in an efficient, robust 
and affine equivariant way. 
A major component of FastMCD is the so-called 
{\it concentration step} (C-step), which works
as follows. 
Given initial estimates $\hbmu_{old}$ for the
center and $\hbSigma_{old}$ for
the scatter matrix, we do:
\begin{enumerate}
\item Compute the distances of all $n$ 
  observations as
	\begin{equation}
		d_{old}(i) = d(\bx_i,\hbmu_{old},
		\hbSigma_{old}).
		\label{eq:cstep1}
	\end{equation}
\item Sort these distances, yielding a permutation 
  $\pi$ for which
	\begin{equation*}
	 d_{old}(\pi(1)) \leqslant d_{old}(\pi(2))
	\leqslant \ldots \leqslant d_{old}(\pi(n)).
		\label{eq:cstep2}
	\end{equation*}
\item Define the $h$-subset $H_{new}$ as
	\begin{equation*}
		H_{new} =
		\left\{\pi(1),\pi(2),\ldots,\pi(h)\right\}.
		\label{eq:cstep3}
	\end{equation*}
\item Compute the new estimates based on $H_{new}$\,:
\begin{align}
 	\hbmu_{new} &=\frac{1}{h}
	 \sum_{i  \,\textrm{in}\,  H_{new}}\bx_i \;,
			\label{eq:cstep4a} \\
	\hbSigma_{new} &=\frac{1}{h-1}
	    \sum_{i \,\textrm{in}\,  H_{new}} 
			(\bx_i-\hbmu_{new})
	    (\bx_i-\hbmu_{new})^T. 
			\label{eq:cstep4b}
\end{align}
\end{enumerate}
Proposition 1 in~\cite{Rousseeuw:FastMCD} 
showed that 
$\textrm{det}(\hbSigma_{new}) \ls 
\textrm{det}(\hbSigma_{old})$, 
with equality if and only if 
$\hbSigma_{new} = \hbSigma_{old}$\,. 
When C-steps are applied iteratively, the 
sequence of determinants must therefore 
converge.

FastMCD starts by drawing a random $(p+1)$-subset 
from the data. Next, its mean and covariance 
matrix serve as $\hbmu_{old}$ and 
$\hbSigma_{old}$ in a C-step. 
The algorithm draws many such $(p+1)$-subsets,
applies several C-steps to each, and keeps the 
solution with the overall lowest determinant.

The computational cost of FastMCD obviously depends 
on $n$ and $p$, but also on the number of random 
$(p+1)$-subsets.  
The default number of initial subsets is  
500, but \cite{Hubert:DetMCD} 
illustrates that this is insufficient at 
high contamination levels when $p$ exceeds 10, 
independent of the sample size $n$. 
In those situations a substantially larger number 
of initial subsets would be required, thereby 
increasing the computational cost significantly.  

\subsection{The DetMCD algorithm}
\label{sec:detmcd}
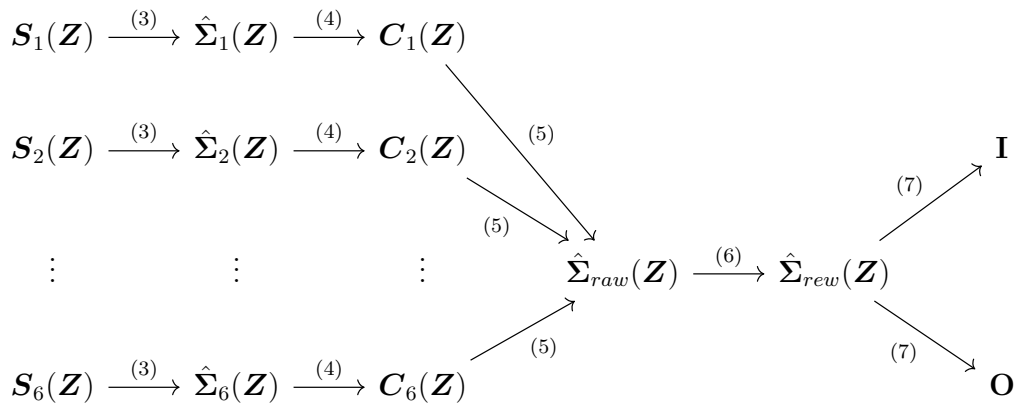
\begin{figure}[hbt]
\centering
\vskip0.2cm
\begin{adjustbox}{width=1.0\textwidth}          
\begin{tikzcd}
	\bS_1(\bZ) \arrow[r,"(3)"] & \hbSigma_1(\bZ)
		\arrow[r,"(4)"] & \bC_1(\bZ) 
		\arrow[rdd,"(5)"] &  &  &  \\
	\bS_2(\bZ) \arrow[r,"(3)"] & \hbSigma_2(\bZ)
		\arrow[r,"(4)"] & \bC_2(\bZ) 
		\arrow[rd,"(5)"'] &  &  & \bf I \\
	\vdots & \vdots  & \vdots  & 
		\hbSigma_{raw}(\bZ) \arrow[r,"(6)"] & 
		\hbSigma_{rew}(\bZ) \arrow[ru,"(7)"] 
		\arrow[rd,"(7)"'] & \\
	\bS_6(\bZ) \arrow[r,"(3)"] & \hbSigma_6(\bZ) 
		\arrow[r,"(4)"] & \bC_6(\bZ) 
		\arrow[ru,"(5)"'] & & & \bf O
\end{tikzcd}
\end{adjustbox}
\vskip0.2cm
\caption{The DetMCD algorithm. 
From left to right: six scatter matrices $\bS_k$ 
from step 2 are refined (step 3) to 
$\hbSigma_k(\bZ)$, followed by C-steps
until convergence (step 4). 
The matrix  $\hbSigma_{raw}(\bZ)$ is the
$\bC_k(\bZ)$ with the lowest determinant 
(step 5). 
Step 6 creates the reweighted estimate 
$\hbSigma_{rew}(\bZ)$ which is then used
to flag outliers (step 7).}
\label{fig:detmcd}
\end{figure}

As an alternative the DetMCD algorithm 
\citep{Hubert:DetMCD} was constructed. 
It is fully deterministic as it does not use 
random subsets. 
It is more robust than FastMCD, and needs 
less computation time. 
The only price to pay is the loss of affine 
equivariance.
DetMCD is only location and scale equivariant, 
but simulations in~\cite{Hubert:DetMCD} showed
that it is very close to affine equivariant. 
The main steps of DetMCD are summarized below, 
and its flowchart is depicted in Figure 
\ref{fig:detmcd}. 
For all details we refer to~\cite{Hubert:DetMCD}. 

\begin{enumerate}
\item Each variable of the dataset $\bX$ is 
  standardized by subtracting its median and 
	dividing by a robust scale estimate, yielding 
	the standardized dataset $\bZ$.
\item Six initial estimates $\bS_k(\bZ)$, 
  $k = 1,\ldots,6$ of the scatter of $\bZ$ are 
	constructed. 
  These initial estimators are fully deterministic 
	and each of them is resistant to certain types 
	of outliers.
\item As the eigenvalues of $\bS_k(\bZ)$ might 
  be inaccurate, they are refined by the routine 
	described in Subsection~\ref{sec:refinement}.
	We denote the resulting covariance 
	matrix by $\hbSigma_k(\bZ)$ and its location
	by $\hbmu_k(\bZ)$.
\item Each $(\hbmu_k(\bZ),\hbSigma_k(\bZ))$ 
  is used to start C-steps
	which are iterated to convergence.
	In each case the resulting scatter matrix is 
	multiplied by $c(\alpha)$ as in 
	\eqref{eq:mcdraw}, yielding the scatter estimate 
	$\bC_k(\bZ)$.
\item The raw DetMCD covariance estimate 
	$\hbSigma_{raw}$ is chosen as the $\bC_k(\bZ)$ 
	with the lowest determinant, with corresponding 
	location estimate $\hbmu_{raw}$.
\item A reweighting step is applied to improve 
  the statistical accuracy as in 
	\cite{Rousseeuw:FastMCD}, 
	yielding the final DetMCD estimates 
	$(\hbmu_{rew},\hbSigma_{rew})$. 
\item The robust distances 
  $\RD_i= d(\bz_i,\hbmu_{rew},
	\hbSigma_{rew})$ then allow to classify the 
	observations into {\bf I}nliers and 
	{\bf O}utliers. 
\end{enumerate}

The DetMCD algorithm thus uses an ensemble of 
initial estimators to ensure high robustness 
against different contamination patterns. 
It is faster than the algorithm in Subsection
\ref{sec:FastMCD}, but not yet fast enough
for real-time applications with high $n$. 
The main bottlenecks are the computation of 
some of the initial estimators $\bS_k$ and 
the time taken by the C-steps. 
The next Section describes how these costs 
can be reduced. 

\section{An improved deterministic MCD}
\label{sec:serial}
\subsection{Standardizing the data}
\label{sec:standardize}
In the first step each variable is 
standardized by means of a robust estimator of 
location and scale. 
Whereas DetMCD used the median and an M-estimator 
of scale, we now use the univariate reweighted MCD 
estimator of \cite{Rousseeuw:Robreg} with 
coverage $\tilde{h} = [n/2]+1$. 
Note that for univariate data, the raw MCD estimates 
reduce to the mean and the standard deviation of 
the $\tilde{h}$-subset with smallest variance. 
They can be computed in $O(n\,\log(n))$ time as 
in \cite{Rousseeuw:Robreg} by sorting the data, 
followed by looping over contiguous 
$\tilde{h}$-subsets while updating their 
means and variances. 
We prefer the univariate MCD because methods that 
give zero-one weights to observations
can be more robust against nearby 
contamination \cite{Raymaekers:DiscMonitoring}. 
The standardized dataset $\bZ$ then consists of
the columns $Z_j = (X_j - \hmu_{uni}(X_j))/
\hsigma_{uni}(X_j)$.  

\subsection{New initial estimators} 
\label{sec:initial}
The six initial estimates used by DetMCD  
are of several types.
The first three estimators start by transforming
the variables one by one, either by
the sigmoid transformation 
$\tilde{Z}_j = \text{tanh}(Z_j)$, 
the rank transformation, or the normal scores 
from the ranks.
The resulting estimator is then the classical 
covariance matrix of the transformed variables.
We will replace these three estimates by a
single new one from \cite{Raymaekers:FROC}, 
using the transformation
\begin{equation}
	\tilde{z}_{ij} = g(z_{ij}) =
	\begin{cases}
	z_{ij} & \mbox{ if } 0 \ls |z_{ij}| \ls b\\
	q_1 \tanh \big(q_2(c-|z_{ij}|)\big) \sign(z_{ij})
	& \mbox{ if } b < |z_{ij}| \ls c \\
	0 & \mbox{ if } |z_{ij}| > c\;.
	\end{cases}
\label{eq:wrapping}
\end{equation}
for $i = 1, \dots, n$  and $j = 1, \dots, p$.
This transformation is called {\it wrapping}.
The default choices are $b=1.5$, $c=4$, 
$q_1 = 1.541$ and $q_2 = 0.862$, which yield 
a continuous function $g$.
These default choices strike a balance 
between accuracy for clean data and robustness 
for contaminated data. 
The choice $b=1.5$ implies that for perfectly
Gaussian data about 85\% of the values are left
unchanged, so that the subsequent computations
remain accurate.
The value $c=4$ reflects that we do not trust 
measurements that lie more than 4 standard 
deviations away.

Next, we compute the new initial estimator
$\btS_1$ as the covariance matrix of the wrapped
data.
In an extensive comparison study 
\cite{Raymaekers:FROC}, this approach was shown 
to perform at least as well as the other
three transformations, so we replace
$\bS_1$, $\bS_2$ and $\bS_3$ by $\btS_1$.

The initial estimators $\bS_4$ and $\bS_5$ in
DetMCD belong to the class of Generalized 
Spatial Sign Covariance Matrices (GSSCM)
\cite{Raymaekers:GenSpatialSign},
which generalizes \cite{Visuri:Rank}.
Among several versions, 
\cite{Raymaekers:GenSpatialSign}
concluded that the so-called
{\it linearly redescending} GSSCM performed 
very well, so we will use it as our second 
initial estimator $\btS_2$.
It is defined as
\begin{equation} 
\btS_2 = \frac{1}{n}\sum_{i=1}^{n}
  {\xi^2(||\bz_i||)\,\bz_i \bz_i^T}
\label{eq:LR}	
\end{equation}
where the weight function $\xi$ is given by
\begin{equation*}
\displaystyle
\xi(r) = 
\begin{cases}
1   & \mbox{ if } r \ls A \\
(B - r)/(B - A) &
\mbox{ if } A < r \ls B \\
0   & \mbox{ if }  r > B \;.
\end{cases}
\end{equation*} 
The cutoffs $A$ and $B$ depend on the set of
norms $||\bz_i||$ as detailed in 
\citep{Raymaekers:FROC}. 
In particular, $A$ is roughly equal to the
median of the $||\bz_i||$.
We replace $\bS_4$ and $\bS_5$ by $\btS_2$,
which achieves a breakdown value of 50\%. 

The final initial estimator $\bS_6$ was the 
OGK estimator~\cite{Maronna:OGK}. 
Whereas $\bS_6$ performed quite well, it was 
by far the most computationally demanding 
among the six initial estimators of DetMCD.
Fortunately simulations showed that the new
$\btS_1$ and $\btS_2$ together are 
sufficient, so we can replace the six initial
estimates by the fast methods $\btS_1$ and 
$\btS_2$ which saves computation time.

\subsection{Refinement of initial estimates} 
\label{sec:refinement}
As our initial estimators $\btS_k$ for $k = 1,2$ 
may have inaccurate or tiny eigenvalues, we 
propose a refinement procedure similar to that in
\cite{Hubert:DetMCD} which uses parts of 
\cite{Maronna:OGK}.
\begin{enumerate}
\item $\btS_k$ is a symmetric matrix so it can be 
	diagonalized as
	\begin{equation*}
		\btS_k = \bV \bD \bV^T
	\end{equation*}
	where $\bV$ is the matrix of eigenvectors of 
	$\btS_k$ and $\bD$ is the diagonal matrix 
	with decreasing eigenvalues 
	$\lambda_1 \gs \dots \gs \lambda_p$. 
	Compute the matrix $\bT$ of principal 
	component scores as 
	\begin{equation*}
		\bT = \bZ \bV\,.
	\end{equation*}
\item If the condition number 
	$\lambda_1/\lambda_p$ of $\btS_k$ exceeds a 
	predefined threshold of (say) 
	$\kappa_{max} = 1000$, then $\btS_k$ is said 
	to be ill-conditioned~\cite{Won:CondNumber}. 
	Then a warning is given and we do not
	continue with $\btS_k$\,.
\item Applying the univariate MCD estimator
	to the scores yields a new diagonal matrix
	\begin{equation*}
		\btD = \text{diag}(\hat{\sigma}_{uni}^2
		    (T_1), \dots, 
				\hat{\sigma}_{uni}^2(T_p))
		\label{eq:dsigma*}
	\end{equation*} 
	from which we compute the refined scatter 
	matrix as
	\begin{equation*}
		\hbSigma_k = \bV \btD \bV^T\;.
	\end{equation*} 
\item The center of $\bZ$ is estimated by sphering
	the data, yielding $\btZ = \hbSigma_k^{-1/2}\bZ$
	with columns $\tilde{Z}_j$ for $j=1,\ldots,p$.
	The univariate MCD estimator for location is then
	applied to each $\tilde{Z}_j$ and the result is
	transformed back, i.e.
	\begin{equation*}
		\hbmu_k(\bZ) = \hbSigma_k^{1/2} \big( 
		\hmu_{uni}(\tilde{Z}_1), \ldots, 
		\hmu_{uni}(\tilde{Z}_p) \big)^T.
		\label{eq:dmu}
	\end{equation*}
\end{enumerate}

\subsection{Speeding up the C-step by Cholesky 
            decomposition}
\label{sec:cholesky}
Starting from both refined estimators $\hbSigma_k$
we then iterate C-steps as in the DetMCD algorithm. 
The main cost of a C-step is the computation of 
the distances \eqref{eq:cstep1} based on the 
inverse of the covariance matrix 
$\hbSigma_{old}$.  
For this we propose to use the Cholesky 
decomposition, i.e. 
\begin{equation*}
	\hbSigma_{old} = \bL \bL^T
\end{equation*}
with $\bL$ a lower triangular $p \times p$ 
matrix. 
We then compute 
$\by_i = \bL^{-1}(\bz_i - \hbmu_{old})$
by forward substitution. It can 
easily be verified that
\begin{equation*}
	d(\bz_i,\hbmu_{old},
	\hbSigma_{old}) = \|\by_i\|\;.
\label{eq:cholesky}
\end{equation*}
We prefer the Cholesky decomposition over
other approaches as it is 
fast and very stable 
numerically~\cite{Lira:Cholesky}. 
It immediately yields the 
determinant by
$\text{det}(\hbSigma_{old}) =
  (\prod_{j=1}^p L_{jj})^2$ with $L_{jj}$ the 
diagonal elements of $\bL$.  

The Cholesky decomposition also allows us to 
monitor the condition number, following 
Algorithms 4.1 and 5.1 in~\cite{Higham:Condition}. 
If
\begin{equation*}
 ||\hbSigma_{old}||_1 \, ||\hbSigma_{old}^{-1}||_1
 \gs \kappa_{max}
\label{eq:choleskymonitor}
\end{equation*}
we approach singularity, and then the C-step is
not taken.
We thus monitor the condition number in two 
different stages of the algorithm: in the 
refinement procedure of $\btS_k$ (Subsection 
\ref{sec:refinement}) and in each C-step.

\subsection{Further speedup by updating}
\label{sec:updating}
To further speed up the C-step, 
we avoid redoing all computations for the new
$h$-subset. 
Let $H_{old}$ be the current $h$-subset, and 
$H_{new}$ the new one obtained by sorting
distances.
We describe the changes in going from $H_{old}$
to $H_{new}$ by an $n$-dimensional vector 
$\bdelta = (\delta_1,\ldots,\delta_n)^T$ in 
which $\delta_i$ in $\{+1, 0, -1\}$ 
indicates whether observation $i$ enters, stays 
in, or leaves $H_{old}$\;. 
Obviously $\sum_i \delta_i = 0$. 
We will use the sum of squares and cross-products
(sscp) matrix
$\bLambda_{old} = (h-1) \hbSigma_{old}$ 
which is the covariance matrix $\hbSigma_{old}$
without denominator.
Initially $\hbmu_{new}= \hbmu_{old}$ and 
$ \bLambda_{new} =  \bLambda_{old}$. 
We then update the center and the sscp matrix 
sequentially~\cite{Bennett:Single-pass,
Hertzog:Pooling,Riani:FSLarge}
as follows.
For each $i$ with $\delta_i \neq 0$:
\begin{enumerate}
\item The total number of observations in the 
  subset is updated:
	\begin{equation*}
	h \leftarrow h + \delta_i \;.
	\label{eq:propcstep1}
	\end{equation*}
\item The center $\hbmu_{new}$ is updated, 
	  and the contribution of $\bz_i$ before and after 
		the update is computed:
	\begin{align*}
	\bu_i &= \bz_i - \hbmu_{new} \\
	\hbmu_{new} &\leftarrow \hbmu_{new} + 
	\frac{\delta_i}{h} \bu_i \label{eq:muUpdate} \\
	\bv_i &= \bz_i - \hbmu_{new} \;.
	\end{align*}
\item Finally the sscp matrix $\bLambda_{new}$ 
  is updated as
	\begin{equation*}
	 \bLambda_{new} \leftarrow  \bLambda_{new} + 
	   \delta_i \bu_i \bv_i^T \;.
	\label{eq:propcstep2}
	\end{equation*}
\end{enumerate}
This one-pass loop replaces 
\eqref{eq:cstep4a} and \eqref{eq:cstep4b} of 
the original C-step procedure, and accounts for 
a noteworthy speedup. 

When $\sum_i |\delta_i|=2$, i.e.\ when only 
two cases are interchanged, it is even faster 
to update the inverse directly. 
From the Sherman-Morrison-Woodbury identity
\begin{equation*}
 (\bA + \bu \bv^T)^{-1} = 
  \bA^{-1} - \frac{\bA^{-1} 
  \bu \bv^T \bA^{-1}}{1+\bv^T \bA^{-1} \bu}
\end{equation*}
we obtain
\begin{equation*} \label{eq:woodbury}
  \left( \bLambda_{new} + 
	\delta_i \bu_i \bv_i^T\right)^{-1} =  
	\bLambda_{new}^{-1} - \frac{\delta_i}{\Delta_i} 
	\left( \bLambda_{new}^{-1} \bu_i \bv_i^T 
	\bLambda_{new}^{-1} \right)
\end{equation*}
with $\Delta_i \coloneqq (1 + \delta_i  \bv_i^T 
\bLambda_{new}^{-1} \bu_i)$. 
Finally, we update the determinant for each 
change in a case $i$ using the identity
\begin{equation*} \label{eq:scaleddetUpdate}
\det \left( \bLambda_{new} + \delta_i \bu_i 
  \bv_i^T \right) = 
	\Delta_i \det (\bLambda_{new}) \;.
\end{equation*}
After the C-steps have converged, 
we multiply 
$\hbSigma_{new} = \bLambda_{new}/(h-1)$ 
by $c(\alpha)$ as in \eqref{eq:mcdraw}. 

\section{Parallel computation and aggregation}
\label{sec:parallel}
\begin{figure}[htb]
\centering
\vskip0.2cm 
\begin{adjustbox}{width=1\textwidth}       
\begin{tikzcd}
	&  & \tilde{\bS}_1(\bZ^{(1)}) \arrow[r, "(2)"] &
	  \hbSigma_1(\bZ^{(1)}) \arrow[r, "(3)"] &
		\bC_1(\bZ^{(1)}) 
		\arrow[rd, "(4)"] &  & \\
	& \bZ^{(1)} \arrow[ru,"(1)"] 
	  \arrow[rd, "(1)"] & 
	  &  &  & \hbSigma^{(1)}_{raw}(\bZ^{(1)}) 
	  \arrow[rdd] &  \\
	&  & \tilde{\bS}_2(\bZ^{(1)}) \arrow[r, "(2)"] &
	  \hbSigma_2(\bZ^{(1)}) \arrow[r, "(3)"] & 
	  \bC_2(\bZ^{(1)}) 
		\arrow[ru, "(4)"] & & \\
	\bX \arrow[ruu] \arrow[rdd] & \bf \vdots &  &  &              
	  & \bf \vdots & {\hbSigma}_{med}(\bZ)\\ 
	& & \tilde{\bS}_1(\bZ^{(q)}) \arrow[r, "(2)"] &
	  \hbSigma_1(\bZ^{(q)}) \arrow[r, "(3)"] & 
	  \bC_1(\bZ^{(q)}) 
		\arrow[rd, "(4)"] & & \\
	& \bZ^{(q)} \arrow[ru, "(1)"]
	  \arrow[rd, "(1)"] &  &  &
		& \hbSigma^{(q)}_{raw}(\bZ^{(q)}) 
		\arrow[ruu] &                       \\
	&  & \tilde{\bS}_2(\bZ^{(q)}) \arrow[r, "(2)"] & 
	  \hbSigma_2(\bZ^{(q)}) \arrow[r, "(3)"] &
		\bC_2(\bZ^{(q)}) 
		\arrow[ru, "(4)"] &  & 
\end{tikzcd}
\end{adjustbox}
\vskip0.2cm
\caption{First part of the parallel processing 
  topology of RT-DetMCD, which computes $q$ raw 
	scatter estimates.}
\label{fig:pdetmcd1}
\end{figure}
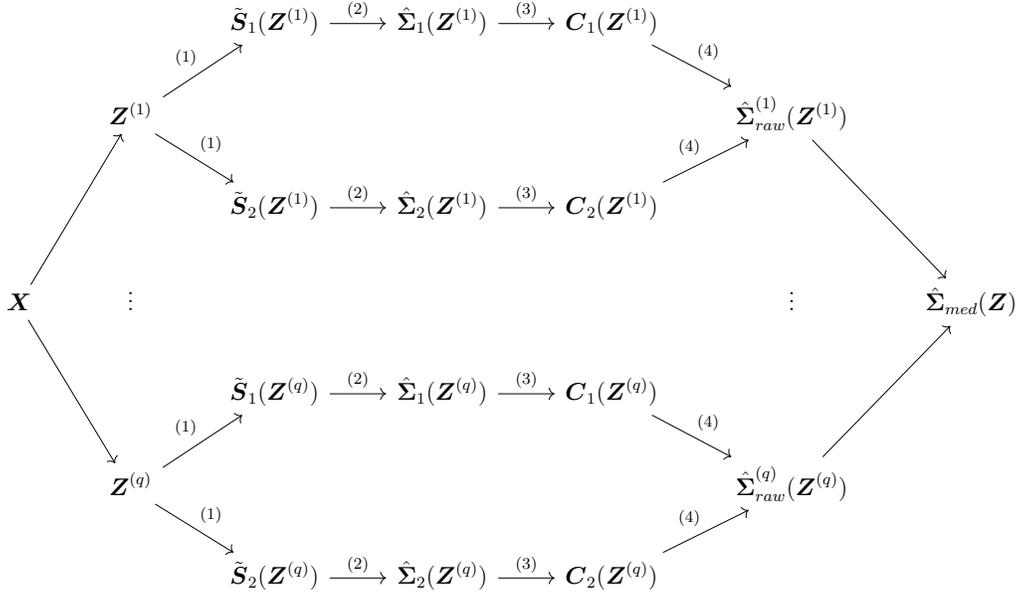

Our final computational improvement stems from 
parallelization. 
Let $\bX$ denote the dataset of $n$ observations 
in $p$ dimensions as before. 
We then randomly partition the dataset in $q$ 
disjoint blocks $\bX^{(l)}$ of $m=\floor{n/q}$ 
cases (discarding the remaining cases if $n$ 
is not divisible by $q$). 
Next, we standardize the blocks by
\begin{equation*}
	\bz_{ij}^{(l)} = \frac{\bx_{ij}^{(l)} - 
	  \hmu_{uni}(X_j)}
		{\hsigma_{uni}(X_j)} 
	\label{eq:standardization}
\end{equation*}
where $l = 1,\dots,q$ and $\hmu_{uni}(.)$ and 
$\hsigma_{uni}(.)$ are the univariate MCD
estimators of location and scale 
(Subsection \ref{sec:standardize}). 
As in Figure \ref{fig:pdetmcd1}
we then use the available processing 
threads as follows.

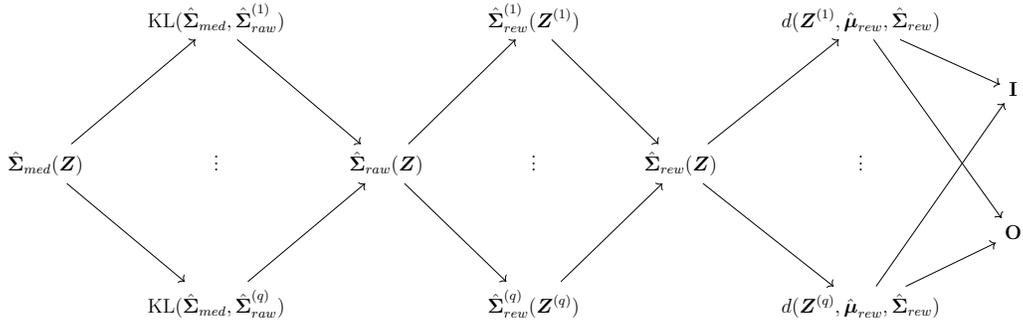
\begin{figure}[htb]
\centering
\vskip0.2cm
\begin{adjustbox}{width=1\textwidth}          
\begin{tikzcd}
  & \mbox{KL}(\hbSigma_{med},\hbSigma^{(1)}_{raw})
	  \arrow[rdd] & & 
		\hbSigma^{(1)}_{rew}(\bZ^{(1)}) \arrow[rdd] & & 
		{d(\bZ^{(1)}, \hbmu_{rew}, \hbSigma_{rew})} 
		\arrow[rd] \arrow[rddd] & \\
  &  &  &  &  &  & \bf I \\
	  \hbSigma_{med}(\bZ) \arrow[ruu] 
	  \arrow[rdd] & \bf \vdots & \hbSigma_{raw}(\bZ) 
	  \arrow[ruu] \arrow[rdd] & \bf \vdots & 
	  \hbSigma_{rew}(\bZ) \arrow[ruu] \arrow[rdd] & 
	  \bf \vdots & \\
  &  &  &  &  &  & \bf O \\
  & \mbox{KL}(\hbSigma_{med},\hbSigma^{(q)}_{raw}) 
	  \arrow[ruu] &  & \hbSigma^{(q)}_{rew}(\bZ^{(q)}) 
		\arrow[ruu] &  & {d(\bZ^{(q)}, 
		\hbmu_{rew}, \hbSigma_{rew})} 
		\arrow[ruuu] \arrow[ru] &      
\end{tikzcd}
\end{adjustbox}
\vskip0.2cm
\caption{Second part of the parallel processing 
topology of RT-DetMCD, responsible for the parallel 
aggregation (left), reweighting (middle) and the 
detection of outliers (right).}
\label{fig:pdetmcd2}
\end{figure}

\begin{enumerate}
\item Compute the initial estimate
  $\btS^{(l)}_1(\bZ^{(l)})$ by wrapping 
	\eqref{eq:wrapping}, and
	$\btS^{(l)}_{2}(\bZ^{(l)})$ by the GSSCM
	method \eqref{eq:LR}.
\item Both estimates are then refined using 
	 the procedure outlined in Subsection 
	\ref{sec:refinement}, which yields 
	$\hbSigma_1(\bZ^{(l)})$ and 
	$\hbSigma_2(\bZ^{(l)})$.
\item We then apply step 4 of the DetMCD 
  algorithm in Subsection \ref{sec:detmcd} to 
	each, using the improvements of Section
	\ref{sec:serial}, yielding
	$\bC_1(\bZ^{(l)})$ and 
	$\bC_2(\bZ^{(l)})$.
\item The raw DetMCD for the block 
  $l=1,\ldots,q$ is then given by
	\begin{equation*}
	(\hbmu^{(l)}_{raw}, \hbSigma^{(l)}_{raw}) 
	  \coloneqq
	\begin{cases}
	(\hbmu^{(l)}_1, \hbSigma^{(l)}_1) & 
	  \text{if } \det(\hbSigma^{(l)}_1) \ls
		\det(\hbSigma^{(l)}_2)\\
	(\hbmu^{(l)}_2, \hbSigma^{(l)}_2) & 
	  \text{otherwise,}
	\end{cases}
	\end{equation*}
	where the type of initial estimator 
	can vary between blocks.
	Note that the percentage of inliers in
	the blocks fluctuates around the
	percentage in the overall dataset, so it 
	is likely that a majority of the $q$ fits
	$(\hbmu^{(l)}_{raw},\hbSigma^{(l)}_{raw})$
	are robust, but some may not be.
\item We now need to aggregate these $q$ 
  fits in a robust way.
  They have many dimensions since the 
	symmetric matrices $\hbSigma^{(l)}_{raw}$
	contain $p(p-1)/2$ distinct entries, and
	the  $\hbmu^{(l)}_{raw}$ have $p$ 
	additional entries.
  Since the total dimension will often
	be higher than $q$, computing a typical
	robust estimate of the $q$ fits is 
	problematic.
  Therefore we compute the entrywise 
	median of the $q$ fits, yielding the
	entrywise median of the $\hbmu^{(l)}$ 
	denoted as
	\begin{equation*}
	  \hbmu_{med} = 
		(\med_l((\hbmu^{(l)}_{raw})_1),\ldots,
		 \med_l((\hbmu^{(l)}_{raw})_p)^T
	\end{equation*}
	and the entrywise median
	of all scatter matrices, given by
	\begin{equation}
	 (\hbSigma_{med})_{jk} =
	 \med_l((\hbSigma^{(l)}_{raw})_{jk})
	\end{equation}
	for $j,k=1,\ldots,p$.
	(Instead of the median also other robust 
	univariate estimators could be used.)
	Note that the matrix $\hbSigma_{med}$ is a 
	robust summary, but it does not have to be
	positive definite.
	Therefore, we cannot use $\hbSigma_{med}$
	as a final aggregated outcome.
\item As a measure of how far the $l$-th fit
  $(\hbmu^{(l)}_{raw},\hbSigma^{(l)}_{raw})$
  is from the entrywise median
	$(\hbmu_{med},\hbSigma_{med})$\,,
  each thread computes the Kullback-Leibler 
	deviation 
	$\mbox{KL}[(\hbmu_{med},\hbSigma_{med}),
	(\hbmu^{(l)}_{raw},\hbSigma^{(l)}_{raw})]$
	given by
	\begin{multline}
	\mbox{KL}[(\ba,\bA),(\bb,\bB)] \coloneqq 
	  \text{trace}(\bA \bB^{-1}) - p
		- \log(\det(\bA \bB^{-1})) \\
		+ (\ba-\bb)^T\bB^{-1}(\ba-\bb)\;.
	\label{eq:KL}
	\end{multline}	
	The quantity 
	$\mbox{KL}[(\ba,\bA),(\bb,\bB)]$ 
  is nonnegative. 
  It is zero when $\ba=\bb$ and $\bA=\bB$, low 
  when $(\ba,\bA)$ deviates little from 
  $(\bb,\bB)$, and high when they are very 
  different.
	
  Note that Formula \eqref{eq:KL} is 
	not symmetric in its arguments, meaning that 
  $\mbox{KL}[(\ba,\bA),(\bb,\bB)]$
  need not be the same as 
  $\mbox{KL}[(\bb,\bB),(\ba,\bA)]$.
  In fact, \eqref{eq:KL} requires $\bB$ to be 
  invertible but does not require $\bA$ to be 
  invertible. 
  This is why we chose the matrix 
  $\hbSigma^{(l)}_{raw}$ for $\bB$ because it 
	is invertible (its determinant is nonzero), 
  whereas the entrywise median matrix 
  $\hbSigma_{med}$ need not be.
\item Sort the deviations from lowest to
  highest and keep the first 
	$\lceil q/2 \rceil$ 
	estimates.
	To simplify notation we pretend that
	these correspond to $l=1,\ldots,
	\lceil q/2 \rceil$.
	These are the block estimates closest to the
	robust summary $\hbSigma_{med}$\,.
	Since the $\hbSigma_{raw}^{(l)}$ are all 
	positive definite we can now aggregate them.
	A simple way would be to average the 
	matrices $\hbSigma_{raw}^{(l)}$ for
	$l=1,\ldots,\lceil q/2 \rceil$ and all the 
	corresponding centers $\hbmu^{(l)}_{raw}$\,.
	
	Instead we can take the union
	of the corresponding $h$-subsets and compute
	its classical mean and covariance matrix.
	A faster way to do this is by a single-pass 
	pooling method \cite{Bennett:Single-pass}.
	We initialize the sscp matrix
	$\bLambda_{pooled}$ by
	$(m-1)\hbSigma_{raw}^{(1)}$ and 
	$\hbmu_{pooled}$ by $\hbmu^{(1)}_{raw}$,
	and set $n_{pooled} = m$.
	Denoting the results from the next
	block by $(\hbmu,\hbSigma)$ we
  \begin{enumerate}
  \item compute the difference in location 
	  $\hbmu_\Delta = \hbmu - \hbmu_{pooled}$ 
		and the sscp matrix 
		$\bLambda = (m-1)\hbSigma$\,.
  \item update the pooled sscp matrix, center 
	  and observation count by
	\begin{equation*} 
	  \bLambda_{pooled} \leftarrow \bLambda_{pooled} + 
		\bLambda + \hbmu_\Delta \hbmu_\Delta^T \, 
		\frac{n_{pooled} \; m}{n_{pooled} + m}\;,
	\label{eq:covpool1}
	\end{equation*}
	\begin{equation*}
		\hbmu_{pooled} \leftarrow \frac{n_{pooled} 
		\; \hbmu_{pooled} + m \; \hbmu}
		{n_{pooled} + m}\;,
	\end{equation*}
	\begin{equation*}
		n_{pooled} \leftarrow n_{pooled} + m\;,
	\label{eq:covpool2}
	\end{equation*}
	\end{enumerate}
	and we continue this way until all
  blocks have been pooled. We then put
	$\hbSigma_{raw}(\bZ) \coloneqq 
	 \bLambda_{pooled}/(n_{pooled}-1)$\,.
\item Next we need to compute the 
  reweighted MCD 
  estimate $(\hbmu_{rew},\hbSigma_{rew})$
	as described in Section \ref{sec:DetMCD}.
	For this we compute the robust distances  
	$\RD_i^{(l)} = d(\bz_i^{(l)},\hbmu_{raw}, 
	\hbSigma_{raw})$ for all blocks $l$ and all 
	cases $i=1,\dots, m$ in each.
	Doing this in the master thread would take
	too long, so we again distribute this 
	computation over the threads.
	Each thread thus obtains a reweighted
	estimate $(\hbmu^{(l)}_{rew},
	\hbSigma^{(l)}_{rew})$.
\item The master thread receives 
	 all local weights and reweighted estimates,  
	 and combines them into the final overall 
	 reweighted estimate 
	 $(\hbmu_{rew},\hbSigma_{rew})$ 
	 by a pooling process similar to step 7 
	 above.
\item Finally, each thread computes robust 
  distances relative to the reweighted estimates 
	and flags the outliers in parallel as those 
	cases whose final robust distance 
	$d(\bz_i^{(l)},\hbmu_{rew},\hbSigma_{rew})$
	exceeds $c_p$\,.
\end{enumerate}
The proposed aggregation strategy is depicted in 
Figure \ref{fig:pdetmcd2}.

Note that the final estimate
$(\hbmu^{(l)}_{rew},\hbSigma^{(l)}_{rew})$
obtained at the end of step 9 can be used
as a ``warm start'' input to step 3 in 
a subsequent run of the algorithm, when 
additional data require updating the result.

\section{Simulations}
\label{sec:sim}
This section analyzes the statistical and 
computational performance of 
RT-DetMCD. 
We proposed three different algorithmic 
modifications in Section \ref{sec:serial}
and one in Section \ref{sec:parallel}.
Switching them on one after the other 
yields the five variations depicted in 
Table \ref{tbl:improvements}. 
The top row is DetMCD without any 
modifications.
The next versions (rows) switch on
modifications: 
new {\bf I}nitial estimators (I), 
{\bf D}istance calculation by Cholesky 
decomposition (D), 
update-based {\bf C}-steps (C), and 
parallelization (P). 
Version IDC is the serial version of 
RT-DetMCD
which does not require a parallel 
architecture.
The parallel version of RT-DetMCD is 
abbreviated as IDCP$_q$ where the subscript 
$q$ denotes the number of blocks used.
Comparing the computation times of the 
different versions is fair, as they
share a common C++ codebase.

\begin{table}[htb]
\centering
\caption{The DetMCD algorithm and four
	 increasingly modified versions.}
\resizebox{1\textwidth}{!}{\begin{tabular}{@{}lllcccc@{}}
\toprule
	Estimator & Section & Remark & \textbf{I}nitial & 
	\textbf{D}istance & \textbf{C}-steps & 
	\textbf{P}arallelization\\ 
\midrule
DetMCD & \ref{sec:DetMCD}  & DetMCD& $\circ$ & $\circ$ &
	$\circ$ & $\circ$\\ 
I & $+$  \ref{sec:standardize}, \ref{sec:initial}, 
	\ref{sec:refinement} & & $\bullet$ & $\circ$ & 
	$\circ$ & $\circ$\\
ID & $+$ \ref{sec:cholesky} & &$\bullet$ & $\bullet$ & 
  $\circ$ & $\circ$\\ 
IDC & $+$ \ref{sec:updating} & Serial RT-DetMCD & 
  $\bullet$ & $\bullet$ & $\bullet$ & $\circ$\\  
IDCP$_q$ & $+$ \ref{sec:parallel} & Parallel RT-DetMCD &
  $\bullet$ & $\bullet$ & $\bullet$ & $\bullet$\\
\bottomrule
\label{tbl:improvements}
\end{tabular}}
\end{table}

We will generate $n$ cases from a $p$-variate 
Gaussian distribution $N(\mathbf{0},\bSigma)$
with center zero, where $p$ is set to 4, 8 or 16 
and $n$ depends on the experiment.
Without loss of generality we set the diagonal
of $\bSigma$ to 1.
Since the methods under consideration are not 
affine equivariant we cannot just set $\bSigma$
equal to the identity matrix.
Instead we consider matrices $\bSigma$ of
different types:
\begin{enumerate}
\item The ALYZ covariance matrices are generated
  as in Section 4 of \cite{Agostinelli:cellwise}, 
	yielding a different $\bSigma$ in each 
	replication. 
	These matrices typically contain
	relatively weak correlations.
\item The A09 type is defined by 
  $\bSigma_{jk} = (-0.9)^{|j-k|}$ for
	$j,k = 1,\ldots,p$.
	This allows for some strong correlations.
\end{enumerate}

Next, we replace $\floor{\eps n}$ random cases 
by outliers of different types, where $\eps$ 
denotes the fraction of contamination. 
{\it Shift contamination} was generated from 
$N(\bmu_C, \bSigma)$ where $\bmu_C$ lies in 
the direction where the outliers are hardest 
to detect, namely that of the last eigenvector 
$\bv$ of the true covariance matrix $\bSigma$.
We rescale $\bv$ to the typical size of
a data point by making 
$\bv^T \bSigma^{-1} \bv = E[Y^2] = p$ where
$Y^2 \sim \chi^2_p$\;.
Finally $\bmu_C = \gamma \bv$ in which $\gamma$
can be varied. 
{\it Cluster contamination} stems from 
$N(\bmu_C, 0.05^2\,\bI)$ where $\bI$ is the 
identity matrix. 
Finally, {\it point contamination} places all 
outliers in the point $\bmu_C$ so they behave
like a tight cluster. 
These settings make the simulation consistent 
with those in \cite{Boudt:MRCD} and
\cite{Hubert:DetMCD}.

The distance of an estimated $\hbSigma$ 
to the true $\bSigma$ is measured 
by the  Kullback-Leibler deviation
$\mbox{KL}(\hbSigma,\bSigma)$ using 
\eqref{eq:KL} without the centers, that is,
\begin{equation*}
  \mbox{KL}(\bA,\bB) =
   \text{trace}(\bA \bB^{-1})
	 - p - \log(\det(\bA \bB^{-1}))\;.
\end{equation*}
This measure was used in several other 
simulation studies such as
\cite{Agostinelli:cellwise, Boudt:MRCD, 
Raymaekers:GenSpatialSign}.
We will compare the accuracy of the new 
methods to that of DetMCD, and also compute 
the speedup factor as
\begin{equation*}
\mbox{speedup} = \mbox{time}(\mbox{DetMCD})/
\mbox{time}(\mbox{new method})\;.
\label{eq:speedup}
\end{equation*} 

The first experiment has $n=2^{16}=65536$ 
observations in $p=4, 8, 16$ dimensions. 
In all versions of MCD we set $\alpha = 0.5$
so $h \approx n/2$ observations are covered, 
which is the most robust choice.
Table \ref{tbl:KL_A09} is for $\bSigma$ of 
type A09 and $\gamma=50$.
The scenarios are point contamination (left), 
shift contamination (middle) and cluster 
contamination (right), both for $10\%$ and 
$30\%$ of outliers.
The top panel presents the KL deviations and
the bottom panel reports the corresponding 
speedup factors, each averaged over 50 
replications.
Table \ref{tbl:KL_ALYZ} shows the same results
for $\bSigma$ of type ALYZ.

\begin{table}[htb]
\caption{Kullback-Leibler deviation and 
	  speedup for $\bSigma$ of type A09.}
\label{tbl:KL_A09}
\centering			
\begin{adjustbox}{width=1\textwidth}					
\centering
\begin{tabular}{@{}lSSScSSScSSS@{}} 
\toprule				
	& \multicolumn{3}{c}{Point contamination} & \phantom{abc}& \multicolumn{3}{c}{Shift contamination} & 
			\phantom{abc} & \multicolumn{3}{c}{Cluster contamination}\\ 
			\cmidrule{2-4} \cmidrule{6-8} \cmidrule{10-12}				
			& {$p=4$} 	& {$p=8$} & {$p=16$} && {$p=4$} & {$p=8$} &  {$p=16$} &&  {$p=4$} &  {$p=8$} &  {$p=16$}\\ 
\\
\multicolumn{9}{@{}l}{\textbf{A: KL deviation}}\\
\midrule				
	$\eps=0.1$ \\			
		DetMCD    		& 	 0.022591  &  0.024252   &  0.026582   &&  0.022693   &  0.024199   &  0.026585   &&  0.022864  &    0.02411   &  0.026596  \\
			I     			& 	 0.022533  &  0.024591   &  0.026621   &&  0.022775   &  0.024361   &  0.026507   &&  0.023043  &   0.024087   &  0.026408  \\
			ID     			& 	 0.022629  &  0.024777   &  0.026649   &&  0.022675   &  0.024471   &  0.026637   &&  0.022976  &   0.024232   &  0.026471  \\
			IDC     		& 	 0.022683  &  0.024845   &  0.026249   &&  0.022658   &  0.024264   &  0.027058   &&  0.023039  &   0.024087   &  0.026578  \\
			IDCP$_{4}$	 	& 	 0.023336  &  0.025806   &  0.028036   &&   0.02325   &  0.025839   &  0.028725   &&   0.02453  &   0.025182   &   0.02795  \\
\\
	$\eps=0.3$ \\	
			DetMCD    		& 	   0.37271   &   0.34738   &   0.33564   &&   0.37339   &   0.34502   &   0.33597   &&   0.37254   &   0.34417   &    0.3357   \\
			I     			& 	   0.37315   &   0.34753   &   0.33617   &&   0.37572   &   0.34502   &   0.33698   &&   0.37324   &    0.3445   &    0.3362   \\
			ID     			&	   0.37315   &   0.34533   &   0.33592   &&   0.37345   &   0.34741   &   0.33648   &&   0.37307   &   0.34428   &   0.33552   \\
			IDC     		& 	   0.37248   &   0.34762   &   0.33596   &&   0.37318   &   0.34502   &   0.33773   &&   0.37254   &   0.34324   &   0.33785   \\
			IDCP$_{4}$		& 	    0.3755   &   0.34948   &   0.34044   &&   0.38955   &   0.35116   &   0.34304   &&   0.37499   &   0.34814   &   0.34098   \\
\\
\multicolumn{9}{@{}l}{\noindent{\textbf{B: Speedup factor}}} \\	
\midrule
	$\eps=0.1$   \\
			I     		&  	 	   90   &    102.21    &   244.22    &&   195.18    
&   215.42    &   221.59    &&   74    &   231.32    &   303.76	 \\
			ID     		&  	       104.44   &    122.82    &   203.12    &&   230.77
    &   272.94    &    268.8    &&   88    &   260.59    &    239.6	 \\
			IDC     	&	       113.2    &   137.38     &  291.18     &&  270.07     
&  290.92     &  325.01     &&  97    &  296.95     &  332.81	 \\
			IDCP$_{4}$	&  	       115.89   &    148.34    &   290.81    &&   357.25
    &   375.96    &   349.84    &&   112.44    &   295.29    &   323.04	 \\
\\
			$\eps=0.3$ \\
			I     		    &	  	335.58    &   419.11   &    432.15   &&     96    &   133.68    &   227.35    &&   119.27    &   284.71    &   296.53 \\
			ID     		 	&       408.45    &   481.35   &    499.55   &&    119.48    &   149.44    &   265.24    &&    145.9    &   311.54    &   325.16  \\
			IDC     	    &	    477.85    &   515.82   &    571.83   &&    126.01    &   161.91    &   297.38    &&   165.95    &   295.81    &   335.75  \\
			IDCP$_{4}$	    &	    573.92    &   557.39   &    687.19   &&    140.24    &   176.72    &   404.91    &&   182.81    &   364.76    &   394.51  \\
			\bottomrule			
		\end{tabular}
	\end{adjustbox}
\end{table}

\begin{table}[htb]
\caption{Kullback-Leibler deviation and 
	  speedup for $\bSigma$ of type ALYZ.}
\label{tbl:KL_ALYZ}
\centering			
\begin{adjustbox}{width=1\textwidth}					
\centering
\begin{tabular}{@{}lSSScSSScSSS@{}} 
\toprule				
		& \multicolumn{3}{c}{Point contamination} & \phantom{abc}& \multicolumn{3
}{c}{Shift contamination} & 
			\phantom{abc} & \multicolumn{3}{c}{Cluster contamination}\\ 
			\cmidrule{2-4} \cmidrule{6-8} \cmidrule{10-12}				
			& {$p=4$} 	& {$p=8$} & {$p=16$} && {$p=4$} & {$p=8$} &  {$p=16$} &&  {$
p=4$} &  {$p=8$} &  {$p=16$}\\ 
\\
\multicolumn{9}{@{}l}{\textbf{A: KL deviation}}\\
\midrule				
	$\eps=0.1$ \\
		DetMCD    		& 	  0.022705  &   0.02424   &  0.026506   &&  0.022697  &   0.024057   &  0.026725   &&  0.022918   &  0.024361  &   0.026383	\\	
			I     			&     0.023027  &   0.02423   &  0.026434   &&  0.023345  &   0.024146   &  0.026881   &&  0.022994   &  0.024442  &   0.026469   	\\
			ID     			&     0.022712  &   0.024697  &  0.026668   &&  0.023319  &   0.025366   &  0.026751   &&  0.022994   &  0.024422  &   0.02633   	\\
			IDC     		&     0.02279   &   0.024231  &  0.027225   &&  0.022855  &   0.024676   &  0.02642    &&  0.022939   &  0.024442  &   0.026398   	\\
			IDCP$_{4}$	 	&     0.023641  &   0.025651  &  0.028605   &&  0.023693  &   0.032569   &  0.029165   &&  0.023693   &  0.02619   &   0.028301   	\\
\\
	$\eps=0.3$ \\	
			DetMCD    		&      0.37209   &   0.34767   &   0.33921   &&   0.37265    &  0.34549   &   0.33606   &&   0.37381   &   0.34677   &   0.33372		\\
			I     			&      0.37216   &   0.34602   &    0.3391   &&   0.37531    &   0.3471   &   0.33616   &&   0.37545   &   0.34751   &   0.33738	\\
			ID     			&	   0.37212   &   0.34534   &    0.3391   &&   0.37283    &  0.34604   &   0.33646   &&   0.37319   &   0.34751   &   0.33515	\\
			IDC     		&      0.37212   &   0.34715   &    0.3391   &&   0.37295    &  0.34487   &   0.33691   &&   0.37319   &   0.34765   &   0.33515	\\
			IDCP$_{4}$		&      0.37512   &   0.35102   &    1.6209   &&   0.37902    &  0.34881   &   0.34296   &&    0.3817   &   0.35386   &   1.0175	\\
\\
\multicolumn{9}{@{}l}{\noindent{\textbf{B: Speedup factor}}} \\
\midrule
	$\eps=0.1$ \\
			I     		&   	83   &   196.54   &    237.67   &&    176.11    &   158.23    &   237.96   &&    202.52    &   227.17    &   182.77  	\\
			ID     		&   	98   &   218.7    &   304.15    &&   210.69     &  190.29     &  240.82    &&   240.71     &  255.71     &  249.31  	\\
			IDC     	&		108.99   &   266.57   &    378.15   &&    269.25    &   213.57    &   306.67   &&    268.17    &   272.02    &   239.06  	\\
			IDCP$_{4}$	&   	 134.4   &   329.91   &    463.72   &&     346.6    &   218.98    &   395.36   &&    324.93    &   360.27    &   348.64  	\\
\\
	$\eps=0.3$ \\
			I     		     &	     255.97    &   263.09   &    281.07    &&   187.99   &    233.34    &   257.78     &&  195.33     &  190.32    &   267.08	\\
			ID     		 	 &       317.78    &   301.18   &    318.24    &&   216.44   &    269.77    &   280.96     &&  238.63     &  222.59    &   297.28	\\
			IDC     	     &	     347.12    &   325.96   &    364.32    &&   241.22   &    285.72    &   315.03     &&  264.22     &  235.65    &   323.08	\\
			IDCP$_{4}$	     &	     473.36    &   356.52   &    378.96    &&   348.88   &     320.4    &    382.5     &&  357.61     &  292.48    &   416.09	\\
\bottomrule
\end{tabular}
\end{adjustbox}
\end{table}

The DetMCD method is in the first row of all 
panels.
The next row contains the I version, which 
modifies the original DetMCD algorithm by
incorporating the new data standardization 
described in Subsection \ref{sec:standardize} 
and replacing the six initial estimators 
by the two new ones of Subsection 
\ref{sec:initial}.
The I version is much faster than the original
DetMCD as seen in its substantial speedup 
factors in both Tables 
\ref{tbl:KL_A09} and \ref{tbl:KL_ALYZ}.
This is due to replacing six initial
estimators (including a slower one) by two 
fast ones.

Note that the accuracy of the I version
(as measured by the KL deviation) is as
good as that of the slower DetMCD.
In some instances with lower $\gamma$
(not shown) the I version was actually
more accurate than DetMCD.
This improvement stems from using 
redescending techniques, which assign zero 
weights to observations that lie far away 
from the majority of data, as in 
\eqref{eq:wrapping} and \eqref{eq:LR}. 
The standardization 
(Subsection \ref{sec:standardize}) and
the refinement procedure (Subsection 
\ref{sec:refinement}) both use the
univariate MCD, and the new initial 
estimators are based on wrapping and the 
linearly redescending GSSCM. 
This makes the proposed algorithm even
more robust against contamination.

The next version (ID) switches on the 
numerically more stable distance 
computation by Cholesky decomposition,
followed by the IDC version which also
incorporates the updating mechanism.
These versions do not change the 
KL deviation much, because both would 
be equivalent to version I if numerical
precision were perfect.
But the new implementations do improve
the speedup factor.
Overall IDC was faster than ID 
which in turn was faster than I, so
each modification has contributed to
the speedup.

When the sample size $n$ is large we
need to speed up the computation even
more.  
This can be achieved by adding the parallel 
computation architecture of Section 
\ref{sec:parallel}, yielding the IDCP version.
Tables \ref{tbl:KL_A09} and \ref{tbl:KL_ALYZ}
show IDCP$_{4}$ which splits up the data
into 4 blocks.
This indeed improves the speedup factor.
However, in some situations (here for
$\eps=0.3$ in Table \ref{tbl:KL_ALYZ})
the speedup is at the expense of a higher
KL deviation, i.e. a loss of accuracy.
This is due to the fact that the blocks
have a lower sample size (here $n/4$), and
for high $p$ (here for $p=16$) there
are not always enough cases per dimension 
to provide an accurate estimate of the
underlying covariance matrix.

We therefore need to choose the number of 
blocks carefully.
Parallelization splits up
the $n \times p$ dataset $\bX$ 
into $q$ blocks $\bX^{(l)}$, each with
$m = \floor{n/q}$ observations.
When choosing $q$ we should take care that
the blocks have enough observations per
dimension to yield accurate estimates,
so we impose
\begin{equation*}
  m/p \gs \omega
\label{eq:omega*}
\end{equation*}
and we will try various choices of
$\omega$, starting from $2^{12} = 4096$.
We only consider values of $q$ that
satisfy this condition, i.e.
$q \ls n/(p\,\omega)$.
In particular, if $n/p < \omega$ we will
not parallelize. 
On the other hand we want to choose $q$
as high as possible to obtain the best
speedup.
Combining these constraints yields the
choice
\begin{equation}
 q = \max\,(\,\floor{\frac{n}{p\,\omega}}
     \,,\,1)\,.
\label{eq:partitionsize}
\end{equation}
When this rule yields $q=1$ we 
use the serial algorithm IDC.
In practice, $q$ is further bounded from
above in terms of the available number of 
CPU cores.

In view of these considerations we carried 
out a new experiment with increasing total 
numbers of observations.
We generated datasets with 
$n = 2^{10},2^{11},\ldots,2^{19}$
with $\bSigma$ of type ALYZ and fraction
$\eps=0.3$ of point contamination
with $\gamma=35$.
We let $\omega$ range from $2^{12}$ to $2^{14}$. 
Table \ref{tbl:parallel} summarizes the 
results, with the same panels for the KL 
deviation and speedup as before. 
The bottom panel shows the number of blocks
$q$ as determined from \eqref{eq:partitionsize},
noting that it is 1 for the smaller sample
sizes $n$. 

\begin{table}[htb]
\caption{Kullback-Leibler deviation and speedup 
	factor for $\bSigma$ of type  ALYZ with 
	fraction $\eps=0.3$ of point 
	contamination, where the number of parallel 
	blocks $q$ is given by 
	\eqref{eq:partitionsize}, for various 
	dataset dimensions and values of $\omega$.} 
\label{tbl:parallel}
\centering			
\begin{adjustbox}{width=1\textwidth}					
\centering
\begin{tabular}{@{}lSSScSSScSSS@{}} 
\toprule				
	 & \multicolumn{3}{c}{$\omega = 2^{12} = 4096$} & \phantom{abc}& \multicolumn{3}{c}{$\omega = 2^{13} = 8192$} & \phantom{abc} & \multicolumn{3}{c}{$\omega = 2^{14} = 16384$}\\ 
\cmidrule{2-4} \cmidrule{6-8} \cmidrule{10-12}				
	{$n$} & {$p=4$} 	& {$p=8$} & {$p=16$} && {$p=4$} & {$p=8$} &  {$p=16$} &&  {$p=4$} &  {$p=8$} &  {$p=16$}\\ 
\\
\multicolumn{9}{@{}l}{\textbf{A: KL deviation}}\\
\midrule
			$2^{10}$    		&        0.37996    & 0.59331     & 0.84716		&&	0.42894   &   0.61663   &   0.77763  &&	   0.44746   &    0.4904    &   0.8792 \\
			$2^{11}$    		&        0.41489   &  0.37776    &  0.57111 	&&	0.39676   &   0.41307   &   0.51507  &&	   0.40545   &   0.41091    &  0.54417 \\
			$2^{12}$    		&        0.32801   &  0.34919    &  0.43299 	&&	0.35214   &   0.37544   &   0.44391  &&	   0.36028   &   0.39317    &  0.44479 \\
			$2^{13}$	 		&        0.36242   &  0.35103    &  0.38566 	&&	0.36837   &   0.33802   &   0.36906  &&	   0.36153   &   0.34559    &  0.36161 \\
			$2^{14}$	 		&        0.35985   &  0.34082    &  0.35838 	&&	0.36244   &   0.34939   &   0.36183  &&	   0.35974   &   0.35924    &  0.35193 \\
			$2^{15}$	 		&        0.37391   &  0.35009    &  0.34588 	&&	0.37452   &   0.34909   &     0.347  &&	   0.38321   &   0.34915    &  0.35389 \\
			$2^{16}$			&        0.36967   &  0.34896    &   0.3453 	&&	0.37717   &   0.34929   &   0.33285  &&	   0.36686   &   0.34383    &  0.34285 \\
			$2^{17}$	 		&        0.36962   &  0.34188    &  0.32901 	&&	0.37271   &   0.3385    &   0.33314  &&	   0.37085   &   0.34256    &  0.33105 \\
			$2^{18}$			&        0.369854  &  0.341957   &  0.331568   	&&	0.370866  &   0.345836  &  0.326351  &&	   0.36967   &   0.345939  &   0.325633  \\
			$2^{19}$			&        0.370744  &  0.344979   &  0.335046  	&&	0.37071   & 0.343914 	&  0.333584  &&	   0.370204  &  0.344144   &  0.33253    \\
\\
\multicolumn{9}{@{}l}{\textbf{B: Speedup factor}}\\
\midrule
			$2^{10}$    		&  6.7533     &  10.826     &  13.855 	&& 		7.5367    &   9.9135    &   14.501 &&	     7.4432    &  10.895    &   13.579 \\
			$2^{11}$    		&   9.1453    &   12.743    &   17.316 	&&	   	10.995    &    12.67    &    16.26 &&	     9.3476    &  13.048    &   16.767 \\
			$2^{12}$    		&   13.781    &   19.272    &   22.901 	&&	   14.863    &   18.789    &   23.176 &&	     14.961    &  18.802    &   23.257 \\
			$2^{13}$	 		&   25.802    &   32.275    &   37.306 	&&	   23.635    &   31.719    &   37.028 &&	     26.861    &  31.683    &   36.402 \\
			$2^{14}$	 		&    49.05    &   63.204    &   72.854 	&&	   47.548    &   61.771    &   66.481 &&	     50.751    &  61.842    &    68.67 \\
			$2^{15}$	 		&   160.39    &    121.8    &   128.97 	&&	   96.779    &   110.18    &   127.52 &&	     93.804    &   121.4    &   124.35 \\
			$2^{16}$			&    489.7    &   386.96    &   228.92 	&&	   301.04    &   203.39    &   233.01 &&	     174.33    &  214.23    &   224.85 \\
			$2^{17}$	 		&   1188.2    &   1056.8    &   769.08 	&&	   837.85    &   715.34    &   396.18 &&	     547.01    &  383.59    &   389.47 \\
			$2^{18}$			&   2489.72   &   2447.27   &   2152.31	&&	   2076.8    &   2011.36  &   1361.63 &&	     1678.18   &  1245.67  &    766.19 \\
			$2^{19}$			&   5020.62   &   5251.24   &   5138.49	&&	   4658.3  &  4731.3   &    3858.81  && 4086.56    &  3674.42  & 2613.72  \\
\\
\multicolumn{9}{@{}l}{\textbf{C: Number of blocks}}\\
\midrule
			$2^{10}$    		&  1   &  1   &  1	&&	   1   &  1   &  1 &&	     1   &  1  &   1 \\
			$2^{11}$    		&  1   &  1   &  1	&&	   1   &  1   &  1 &&	     1   &  1  &   1 \\
			$2^{12}$    		&  1   &  1   &  1	&&	   1   &  1   &  1 &&	     1   &  1  &   1 \\
			$2^{13}$	 		&  1   &  1   &  1	&&	   1   &  1   &  1 &&	     1   &  1  &   1 \\
			$2^{14}$	 		&  1   &  1   &  1	&&	   1   &  1   &  1 &&	     1   &  1  &   1 \\
			$2^{15}$	 		&  2   &  1   &  1	&&	   1   &  1   &  1 &&	     1   &  1  &   1 \\
			$2^{16}$			&  4   &  2   &  1	&&	   2   &  1   &  1 &&	     1   &  1  &   1 \\
			$2^{17}$	 		&  8   &  4   &  2	&&	   4   &  2   &  1 &&	     2   &  1  &   1 \\
			$2^{18}$			&  16   &  8   &  4	&&	   8   &  4   &  2 &&	     4   &  2  &   1 \\
			$2^{19}$			&  32   & 16   &  8	&&	  16   &  8   &  4 &&	     8   &  4  &   2 \\
\bottomrule
\end{tabular}%
\end{adjustbox}
\end{table}

In Table \ref{tbl:parallel} we see 
that the KL deviation remained stable over 
all dataset sizes. 
This indicates that provided
$q$ is chosen by \eqref{eq:partitionsize},
i.e. the blocks have enough observations
per dimension,
the accuracy of parallel RT-DetMCD is 
comparable to that of the serial version.
At the same time the parallel version
achieves much higher speedup factors than
the serial version. 
We also note that the estimation accuracy 
was rather stable across the three values 
of $\omega$ considered.
It thus appears that $\omega = 2^{12}$ 
(which yields the best speedup factors) is 
a reasonable default choice.

\section{Industrial application of RT-DetMCD} 
\label{sec:industrialexample}

Industrial food inspection machines scan 
millions of individual objects per hour, 
yielding faster and more accurate results
than manual inspection.
Mechanical sorting boosts the processing 
capacity of a production line, enabling the 
food producer to simultaneously provide 
consistent food quality and safety guarantees. 
We illustrate the feasibility of anomaly 
detection by RT-DetMCD in this context. 
The example is an almond inspection setting, 
where the machine measures the object response 
on $p=4$ wavelengths using a line scan image 
acquisition system. 
Each incoming scan line consists of 4096 pixels 
and has to be classified within milliseconds 
to comply with the production throughput. 
The goal is the adequate detection of foreign 
material (such as shells, hulls, wood, stones
and pieces of glass) between the almonds, 
so the foreign material can be removed in 
real time.

We use the RT-DetMCD method for 
unsupervised classification.
This is considerably different from the 
customary classification setting, where 
training sets from each individual product 
must first be analyzed carefully by hand in 
order to assign its objects to different 
types of material.
Instead, we assume that the training sets are 
contaminated by defects, that is, outliers.

\begin{figure}[htb]
\centering
\includegraphics[width=0.75\linewidth]
  {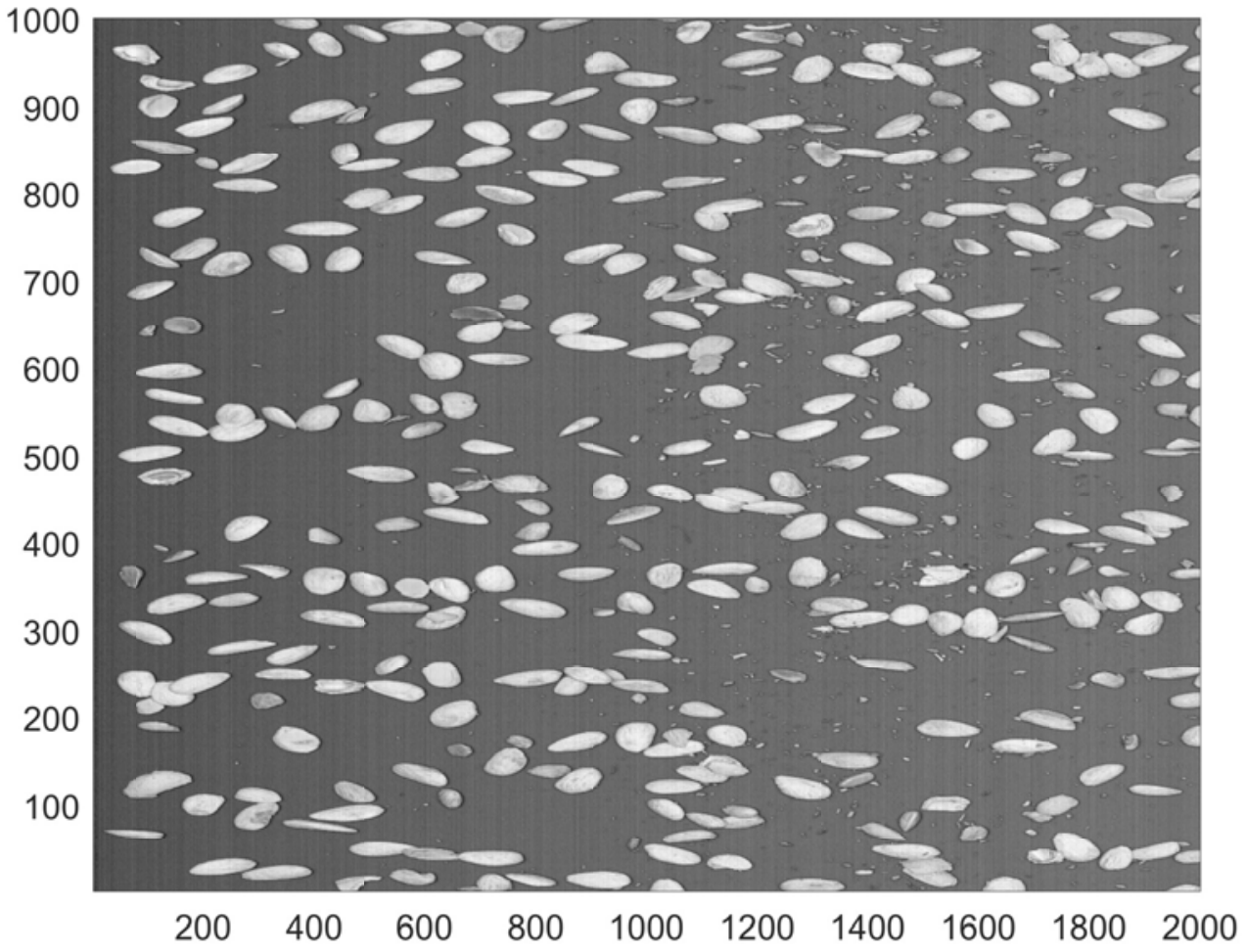}
\caption{$1000 \times 2000$ pixel region of the 
classifier training set. The image contains 
almonds as well as almond shells and dust.}
\label{fig:almondinputimage}
\end{figure}

In the example the training set consists of 
$2048$ sequentially stacked scan lines of 
$4096$ pixels which captured the incoming 
product flow, totaling over 8 million 
observations (pixels) with $p=4$ dimensions
each.
The first dimension of the dataset is 
visualized in black and white in 
Figure \ref{fig:almondinputimage}.
All the images of this example were clipped to 
a region of interest of $1000 \times 2000$ 
pixels so the image resolution can be rendered
here.

We first extract the relevant foreground 
objects by training parallel\linebreak 
RT-DetMCD on all eight million observations, 
yielding a fit $(\hbmu_1,\hbSigma_1)$.
As the majority of these observations consist 
of background (i.e. the dark pixels in Figure 
\ref{fig:almondinputimage}), RT-DetMCD
identified the foreground material as anomalies, 
shown in Figure 5a. 
Next, RT-DetMCD was trained on the 
$3\,127\,973$ foreground objects, 
yielding a fit $(\hbmu_2,\hbSigma_2)$ in 
seconds, which revealed non-almond material 
(Figure 5b). 
Closer inspection showed that entire shells 
were adequately detected as outliers, as well
as almond discolorations and damaged almond 
skins.

The next task was to classify a variety of 
unknown material in a test dataset, i.e. a
previously unseen image of material. 
This was achieved by computing robust 
distances of new observations from the existing 
fit, and checking when they exceed the cutoff.
The computation was done in parallel, using 
the third part of the flowchart in Figure
\ref{fig:pdetmcd2} corresponding to
step 10 in the algorithm in Section
\ref{sec:parallel}.
This construction forms an anomaly detector 
that uses the fits trained on the image shown 
in Figure \ref{fig:almondinputimage}. 
The robust distances from the background 
segmentation fit $(\hbmu_1,\hbSigma_1)$ 
performed as expected, detecting all foreground 
material on the fly (Figure 5c).
It also revealed the presence of water droplets 
on the image acquisition lens, seen as vertical 
stripes around columns 800 and 1000. 
Presented with the foreground objects, the 
second detector based on $(\hbmu_2,\hbSigma_2)$
revealed all non-almond material (e.g. almond 
tree wood), with the output shown in 
Figure 5d. 

\begin{figure}[htb]
\centering
\includegraphics[width=0.9\linewidth]
  {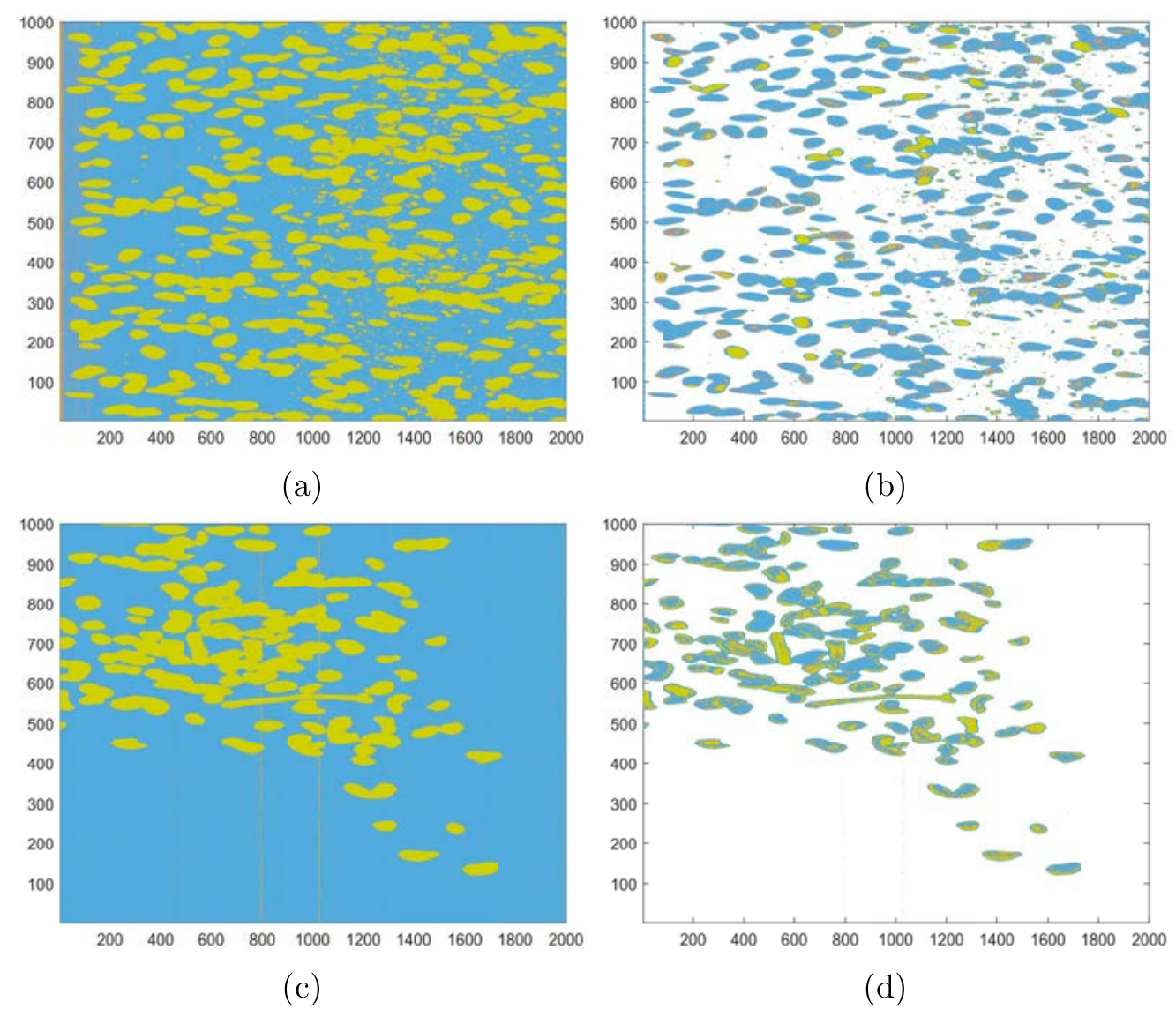}
\caption{Industrial almond dataset: (a) segmenting
the training dataset of Figure 
\ref{fig:almondinputimage} into foreground and 
background by RT-DetMCD with foreground shown 
in yellow; (b) detecting outliers among the 
foreground pixels reveals foreign material 
shown in yellow; correctly detected foreground (c) 
and defects (d) in a test dataset.}
\end{figure}

Segmenting the entire new image (the test
dataset) with over 8 million observations into
background and foreground only took $8.4$ 
milliseconds, whereas segmenting the
approximately 3 million foreground cases
took $3.3$ milliseconds.

Note that in industrial settings the
computation speed of RT-DetMCD is an important
advantage since it means that the classifier
can be re-trained quickly, even on-the-fly
whenever new data are observed.
In this particular application it was sufficient
to run RT-DetMCD at regular intervals. 

\section{Conclusions and outlook}
\label{sec:concl}
Real-time industrial processes are very 
demanding in terms of computation speed.
Often the detection of anomalies is of crucial
importance, e.g. for food sorting machines that
need to remove foreign material on the fly.
This paper focused on anomaly detection by
robust estimation using the minimum covariance
determinant (MCD) approach.

Although the existing DetMCD algorithm is
fast enough for off-line statistical analysis,
it cannot cope with the huge sample sizes and
stringent speed requirements of industrial 
processes.
Therefore we constructed an improved method
called RT-DetMCD by incorporating several new 
ideas, resulting in high speedup factors
without loss of accuracy.
A major speedup is obtained by parallel
processing, which splits up the data into
blocks that are analyzed separately. 
Combining these results into an overall fit
required the development of a novel aggregation
approach. 

The performance of RT-DetMCD was studied by
simulation, which showed that each improvement
contributed to the overall speedup.
Its ability to handle real-time industrial
processes was illustrated by a case study
on the automated sorting of almonds.
The industrial C++ code of RT-DetMCD
used in the simulation and application is
proprietary, 
but a research-level Matlab version which 
mimics its results is available from the 
webpage\linebreak
{\it http://wis.kuleuven.be/statdatascience/robust/software}\,.

The output of the new RT-DetMCD technique 
can be used as a basis for other multivariate 
techniques such as robust principal component 
analysis and classification in industrial 
settings.

\section*{Acknowledgements}
We thank Johan Speybrouck for providing the 
industrial datasets and Tim Wynants for his 
support throughout the project. 
We also acknowledge the financial support of 
VLAIO grant HBC.2016.0208 as well as project
C16/15/068 of Internal Funds KU Leuven.

\newpage


\end{document}